\documentclass[aps,prl,twocolumn,showpacs,psfig,superscriptaddress,longbibliography]{revtex4-1}
\usepackage{amsfonts}
\usepackage{mathrsfs}
\usepackage{amsmath}
\usepackage{color}
\usepackage{natbib}
\usepackage{textcomp}
\usepackage{graphicx}
\usepackage{bm}
\usepackage{amssymb}
\usepackage{xspace}
\usepackage{epstopdf}
\usepackage{dcolumn}
\usepackage{longtable}
\usepackage{multirow}
\usepackage[colorlinks=true, letterpaper=true, pdfstartview=FitV, linkcolor=blue, citecolor=blue, urlcolor=blue]{hyperref}
\usepackage{float}

\makeatletter

\newcommand{\Rmnum}[1]{\expandafter\@slowromancap\romannumeral #1@}
\makeatother
\begin{document}

\title{Two-Dimensional Room Temperature Ferromagnetic Semiconductors with Quantum Anomalous Hall Effect}

\author{Jing-Yang You}
\email{These authors contributed equally to this work.}
\affiliation{School of Physical Sciences, University of Chinese Academy of Sciences, Beijng 100049, China}

\author{Zhen Zhang}
\email{These authors contributed equally to this work.}
\affiliation{School of Physical Sciences, University of Chinese Academy of Sciences, Beijng 100049, China}

 \author{Bo Gu}
 \email{gubo@ucas.ac.cn}
 \affiliation{Kavli Institute for Theoretical Sciences, and CAS Center for Excellence in Topological Quantum Computation, University of Chinese Academy of Sciences, Beijng 100190, China}
\affiliation{Physical Science Laboratory, Huairou National Comprehensive Science Center, Beijing 101400, China}

\author{Gang Su}
\email{gsu@ucas.ac.cn}
\affiliation{School of Physical Sciences, University of Chinese Academy of Sciences, Beijng 100049, China}
\affiliation{Kavli Institute for Theoretical Sciences, and CAS Center for Excellence in Topological Quantum Computation, University of Chinese Academy of Sciences, Beijng 100190, China}
\affiliation{Physical Science Laboratory, Huairou National Comprehensive Science Center, Beijing 101400, China}

\begin{abstract}
To obtain room temperature ferromagnetic semiconductors and
to realize room temperature quantum anomalous Hall effect (QAHE) have been big challenges for a long time. Here we report that, based on first-principles calculations, PdBr$_3$, PtBr$_3$, PdI$_3$ and PtI$_3$ monolayers are ferromagnetic semiconductors that could exhibit high temperature QAHE. Curie temperatures estimated by Monte Carlo simulations are 350 K and 375 K for PdBr$_3$ and PtBr$_3$ monolayers, respectively. The band gaps of PdBr$_3$ and PtBr$_3$ are disclosed to be 58.7 meV and 28.1 meV with GGA and 100.8 meV and 45 meV with HSE06, respectively, being quite well in favor of observing room temperature QAHE. It is shown that the large band gaps are induced from multi-orbital electron correlations. By carefully studying the stabilities of the above four monolayers, we unveil that they could be feasible in experiment. The present work sheds new light on developing spintronic devices by using room temperature ferromagnetic semiconductors, and implementing dissipationless devices by applying room temperature QAHE.
\end{abstract}
\maketitle


{\color{blue}{\em Introduction}}---The realization of materials that combine semiconducting behavior and magnetism has long been a dream of material physics. Magnetic semiconductors have been demonstrated to work at low temperatures but not yet at high temperatures for spintronic applications. ``Is it possible to create magnetic semiconductors that work at room temperature?" has been selected as one of 125 big questions in Science~\cite{Kennedy2005}. The highest Curie temperature of the most extensively studied magnetic semiconductor (Ga,Mn)As is still far below room temperature ~\cite{Dietl2014}. Recent development of magnetism in two-dimensional (2D) van der Waals materials has provided a new class of magnetic semiconductors with possible high Curie temperatures ~\cite{Burch2018}.

In 2D systems, magnetism with nontrivial topology can induce unusual behaviors. One example is the quantum anomalous Hall effect (QAHE), which is a quantized Hall effect without an external magnetic field, where the combination of strong spin-orbit coupling (SOC) and ferromagnetic ordering can generate a band gap in the bulk and gapless chiral edge states at boundaries, and the quantized Hall conductivity is carried by the edge states ~\cite{Haldane1988,Onoda2003,Liu2008,He2018,Liu2016,Kou2015}. Owing to the dissipationless chiral edge states, QAHE would have potential applications in low-power consumption spintronic devices~\cite{Wu2014a}. Thus, the search for materials with QAHE has attracted extensive interests ~\cite{Liu2008,Wu2008,Yu2010,Chang2013,Chang2013a,Si2017}. Inspired by the seminal work of Haldane~\cite{Haldane1988}, the honeycomb lattice is often thought to be a proper platform for QAHE. Graphene with enhanced SOC, which can be topological insulators (TI)~\cite{Kane2005,Weeks2011,Gmitra2016}, may reveal QAHE by doping magnetic impurities or through proximity effect~\cite{Zhang2015,Qiao2014,Zhang2012,Zhang2013,Wu2014}.

Nonetheless, the QAHE in current experiments are only realized at very low temperature. The first experimental observation of QAHE was realized in Cr-doped (Bi,Sb)$_2$Te$_3$ thin film at 30 mK ~\cite{Chang2013}. Later, QAHE was observed in V-doped (Bi,Sb)$_2$Te$_3$ thin film at 25 mK ~\cite{Chang2015} and Cr/V-codoped (Bi,Sb)$_2$Te$_3$ system about 300 mK~\cite{Ou2017}. So far, the highest temperature to implement QAHE is
about 2K in Cr doped (Bi,Sb)$_2$Te$_3$ films~\cite{Mogi2015}. In those doped TIs, magnetic disorder was found to affect dramatically the temperature in observing QAHE. The edge states are robust against lattice disorder, but greatly affected by magnetic disorder. To obtain a full quantization of QAHE in magnetically doped TIs, a very low temperature was usually employed to suppress magnetic disorders in experiments \cite{He2018}.

To overcome the influence of magnetic disorder, a 2D intrinsic ferromagnetic semiconductor with finite Chern number becomes a promising way to implement QAHE. The QAHE with in-plane magnetization was discussed~\cite{Liu2013}, and the QAHE at temperature of about 20 K has been recently proposed in monolayer LaCl with in-plane magnetization ~\cite{Liu2018a}. For realistic applications of QAHE, new materials with room temperature QAHE are highly desired. Several ferromagnetic transition metal trihalides were proposed to be candidates for the implementation of QAHE~\cite{He2016,Huang2017,He2017,Sun2018,Wang2018}. Among them the PdCl$_3$ monolayer was predicted to be a promising candidate for realizing QAHE with high temperature.

In order to find QAHE materials more feasible in experiment, we notice that the bulk PtBr$_3$ was synthesized and described by $W\ddot{o}hler$ in 1925~\cite{Woehler1925} and the single crystals of the dark-red PtBr$_3$ were grown in 1969~\cite{Thiele1969}. Here, we show that, by first-principles calculations, PdBr$_3$ and PtBr$_3$ monolayers with honeycomb lattice are 2D ferromagnetic semiconductors with out-of-plane magnetization, which can also implement the QAHE at quite high temperature, where the Curie temperature $T_C$ and band gap $E_g/k_B$ are shown higher than room temperature. The large band gap, opened by the spin-orbit coupling, comes from the multi-orbital electron correlations. By our calculations, PtBr$_3$ and PdBr$_3$ monolayers are shown to have free energies and formation energies both lower than the recently proposed PdCl$_3$ monolayer~\cite{Wang2018}, indicating that the 2D PdBr$_3$ and PtBr$_3$ may be promising candidates for implementing room temperature QAHE in experiment.

The first-principles calculations were done with the Vienna ab initio simulation package (VASP) using the projector augmented wave (PAW) method in the framework of density functional theory (DFT)~\cite{Kresse1993,Kresse1996}. The electron exchange-correlation functional was described by the generalized gradient approximation (GGA) in the form proposed by Perdew, Burke, and Ernzerhof (PBE) ~\cite{Perdew1996}. The structure relaxation considering both the atomic positions and lattice vectors was performed by the conjugate gradient (CG) scheme until the maximum force on each atom was less than 0.0001 eV/${\AA}$, and the total energy was converged to 10$^{-8}$ eV with Gaussian smearing method. To avoid unnecessary interactions between the monolayer and its periodic images, the vacuum layer is set to 15 {\AA}. The energy cutoff of the plane waves was chosen as 550 eV. The Brillouin zone (BZ)integration was sampled by using a $13\times13\times1$ G-centered Monkhorst-Pack grid for the calculations of relaxation and electronic structures. The phonon frequencies were calculated using a finite displacement approach as implemented in the PHONOPY code ~\cite{Togo2015}, in which a $3\times3\times1$ supercell and a displacement of 0.01{\AA} from the equilibrium atomic positions are employed. SOC is included by a second variational procedure on a fully self-consistent basis. An effective tight-binding Hamiltonian constructed from the maximally localized Wannier functions (MLWF) was used to investigate the surface states ~\cite{Mostofi2014,Kong2017}. The iterative Green function method ~\cite{Sancho1985,Sancho1985} was used with the package WannierTools ~\cite{Wu2018}.

\begin{figure}[!!!h]
  \centering
  \includegraphics[scale=0.37,angle=0]{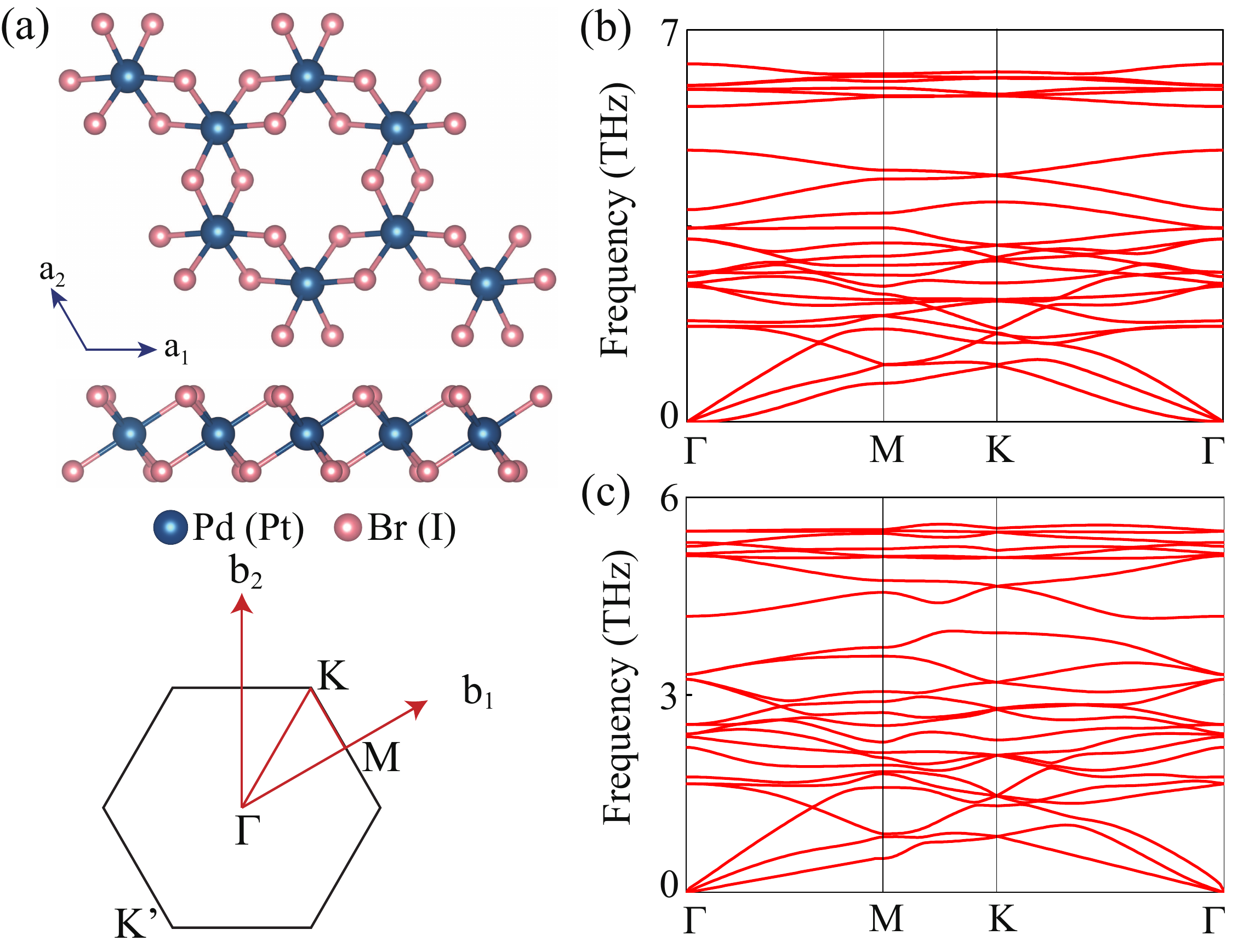}\\
  \caption{{\bfseries Stable structures of 2D PdBr$_3$ and PtBr$_3$ monolayers with honeycomb lattice.} (a) Top and side views of the PdBr$_3$ (PtBr$_3$) monolayer as well as first Brillouin zone with high symmetry points labeled. Pd (Pt) atoms form a honeycomb lattice, and a unit cell contains four Pd (Pt) atoms. Phonon spectra of (b) PdBr$_3$ and (c) PtBr$_3$ monolayers.}\label{fig1}
\end{figure}

{\color{blue}{\em Results}}---The structure of PdBr$_3$ (PtBr$_3$) monolayer is depicted in Fig.~\ref{fig1} (a), whose space group is $P\bar{3}$1$m$ (No.162). Each primitive cell contains two formula units of PdBr$_3$ (PtBr$_3$) and has two Pd (Pt) atoms. Under the crystal field of the surrounding Br octahedra, the $d$ orbitals of the two Pd (Pt) atoms split into threefold $t_{2g}$ orbitals and twofold $e_g$ orbitals, where the latter is energetically higher. For Pd$^{3+}$ (Pt$^{3+}$) with seven electrons in PdBr$_3$ (PtBr$_3$) monolayer, since the crystal field is strong, six electrons will fill $t_{2g}$ orbitals and the remaining one will fill $e_g$ orbitals, leading to the spin S = 1/2. To confirm the stability of PdBr$_3$ (PtBr$_3$) monolayer, its phonon spectra have been calculated. There is no imaginary frequency mode in the whole Brillouin zone as shown in Figs.~\ref{fig1} (b) and (c), indicating that the PdBr$_3$ and PtBr$_3$ monolayers are dynamically stable.

\begin{table}[t]
	\caption{The formation energy $E_f$ for monolayers PdBr$_3$, PtBr$_3$, PdI$_3$ and PtI$_3$ (in eV), as well as PdCl$_3$ monolayer for comparison, calculated by $E_f = E(AB_3)-E(A)-(3/2) E(B_2)$, where $E(AB_3)$ and $E(A)$ are the total energies of the PdBr$_3$ (PtBr$_3$, PdI$_3$ and PtI$_3$) and Pd (Pt) crystals, respectively, while $E(B_2)$ is the total energy of Br$_2$ (I$_2$) molecule.}\label{tab:formation}
	\begin{tabular}{cccccccc}
		\hline
		                     & PdBr$_3$ & PtBr$_3$ &PdI$_3$ &PtI$_3$& PdCl$_3$ \\
		\hline \hline
		E$_f$                & -0.08     &-2.41     &0.13   &-2.44   &0.01    \\
        \hline
	\end{tabular}
\end{table}

\begin{figure}[!hbt]
  \centering
  \includegraphics[scale=0.35,angle=0]{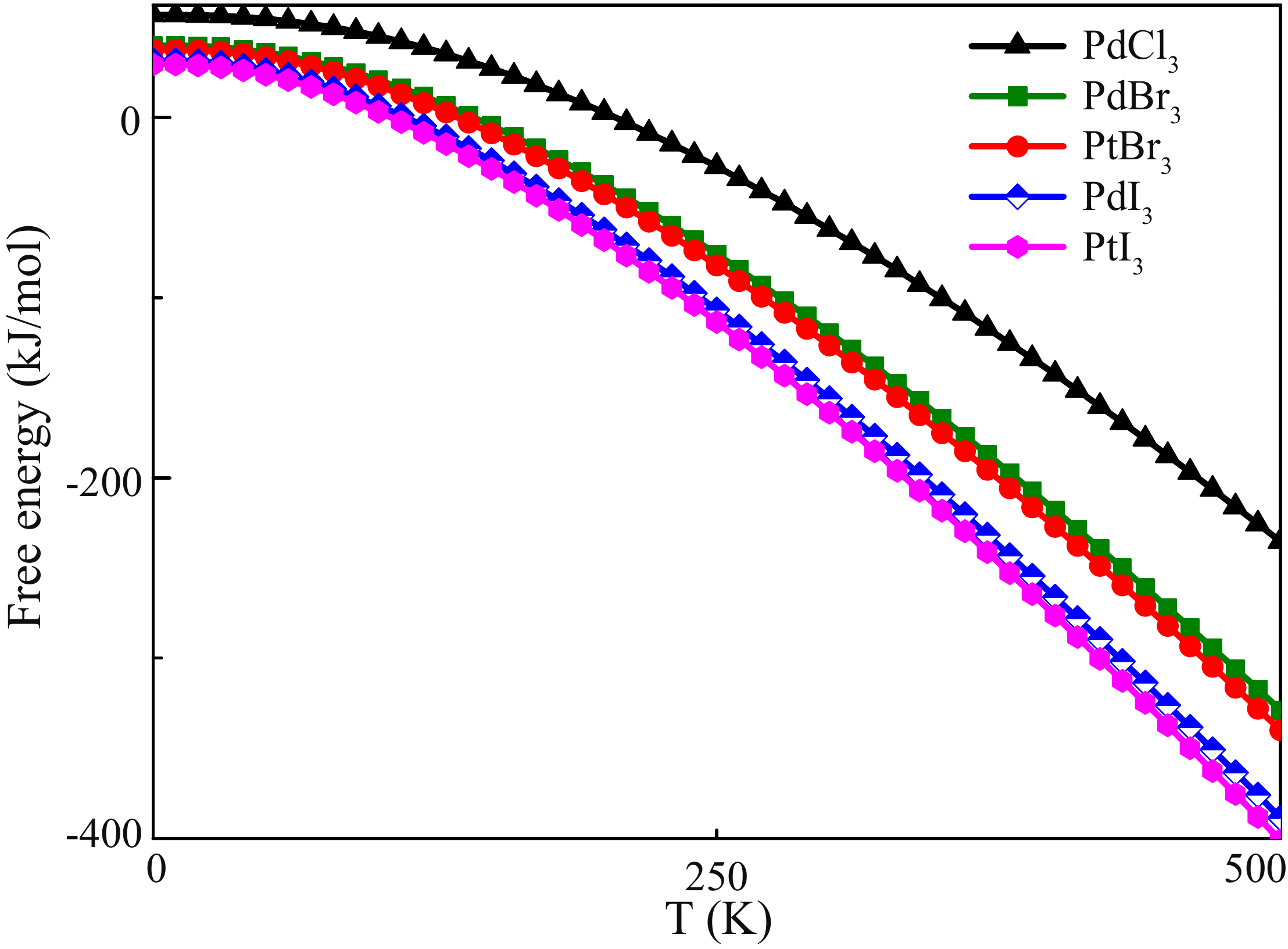}\\
  \caption{The calculated temperature dependent free energy of PdBr$_3$, PtBr$_3$, PdI$_3$ and PtI$_3$ monolayers, where PdCl$_3$ is included for comparison.}\label{fig2}
\end{figure}

The structural stabilities of PdBr$_3$ and PtBr$_3$ are also examined by the formation energy (Table~\ref{tab:formation}) and the free energy (Fig.~\ref{fig2}). The obtained negative values, $E_f $= -0.08 eV and -2.41 eV for PdBr$_3$ and PtBr$_3$ monolayers, respectively, are indicative of an exothermic reaction. We also calculate the formation energy with value of 0.01 eV for PdCl$_3$ monolayer using the same method as PdBr$_3$ and PtBr$_3$, which can thus be compared reasonably at the same level. The formation energy and free energy for PdBr$_3$ and PtBr$_3$ monolayers lower than that for PdCl$_3$, indicates that they are more feasible in experiment.

\begin{figure}[!hbt]
  \centering
  \includegraphics[scale=0.35,angle=0]{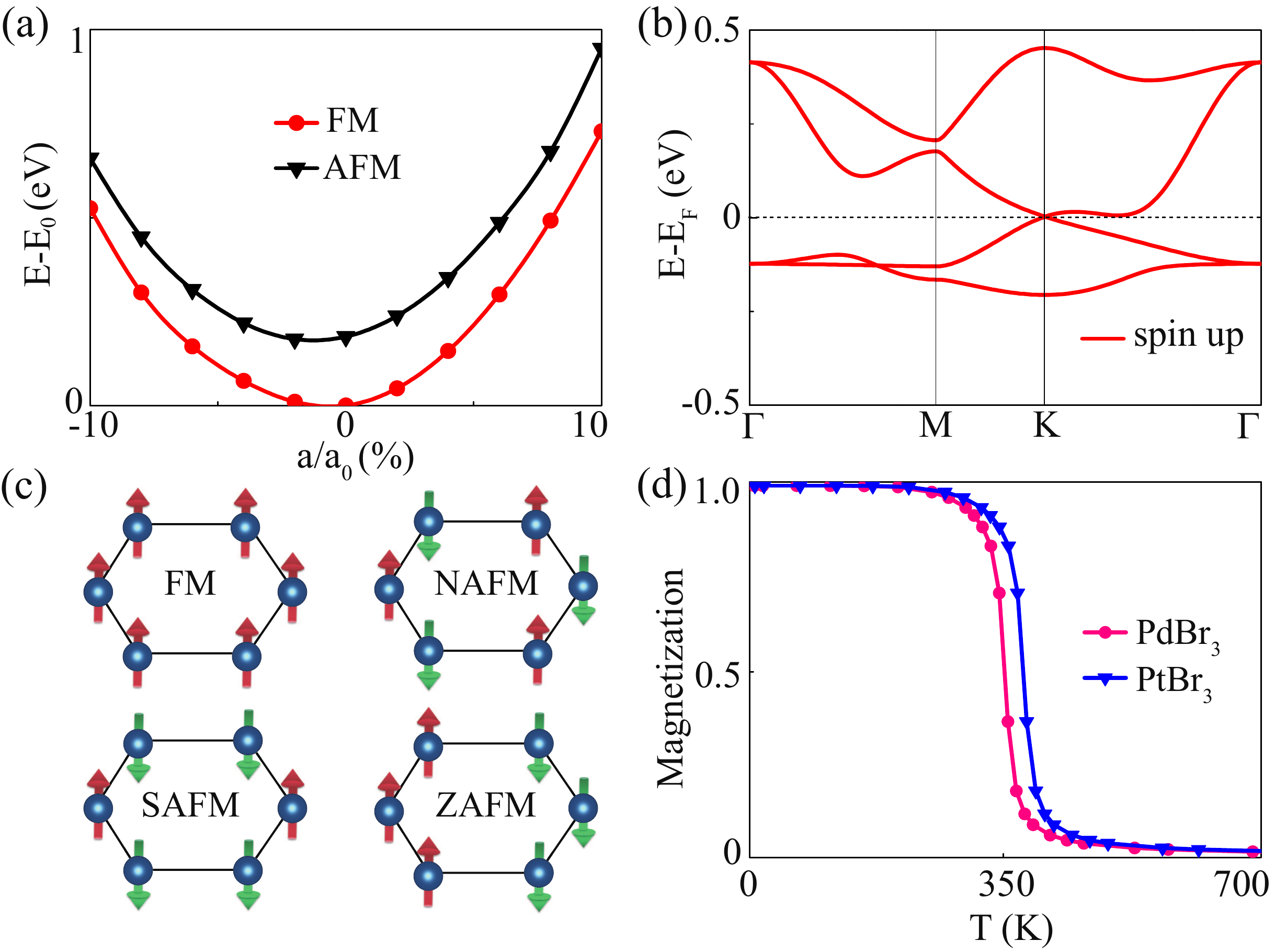}\\
  \caption{{\bfseries Half-ferromagnetic PdBr$_3$ and PtBr$_3$ monolayers with Curie temperatures above room temperature.} (a) Total energy of PdBr$_3$ monolayer as a function of lattice constant for ferromagnetic (FM) and antiferromagnetic (AFM) configurations. (b) Band structure of PdBr$_3$ monolayer without inclusion of the spin-orbit coupling (SOC), where the bands for spin down electrons are too far from the Fermi level to be included.  (c) Possible configurations of Pd (Pt) spins on honeycomb lattice: FM, N$\acute{e}$el AFM (NAFM), stripe AFM (SAFM), and zigzag AFM (ZAFM). (d) Temperature dependence of the normalized magnetic moment of PdBr$_3$ and PtBr$_3$ monolayers by Monte Carlo simulations.}\label{fig3}
\end{figure}

To determine the ground state of PdBr$_3$ monolayer, in the absence of SOC, we calculated the total energy for ferromagnetic (FM) and antiferromagnetic (AFM) configurations as a function of lattice constant, and found that the FM state has an energy lower than AFM state. The equilibrium lattice constant of PdBr$_3$ for FM state is a$_0$ = 6.667 {\AA} in Fig. ~\ref{fig3} (a). The monolayer PtBr$_3$ also possesses the ferromagnetic ground state, but with a larger lattice constant a$_0$ = 6.732 {\AA}.

The electronic band structure without inclusion of SOC [Fig. ~\ref{fig3} (b)] shows that PdBr$_3$ monolayer is a Weyl half-metal, which acts as a conductor with only one species of electron spin orientations at Fermi level and at the high-symmetry point K there exists a Weyl node protected by $C_{3v}$ space group. Considering the inversion symmetry, there is another Weyl point at K$^\prime$. The total density of states (DOS) are mainly contributed by $p$ electrons of Br and $e_g$ electrons of Pd atoms.

In the presence of SOC, the magnetic anisotropy should be considered. To determine the magnetic behavior of PdBr$_3$ and PtBr$_3$ monolayers, we calculated the total energy of PdBr$_3$ and PtBr$_3$ monolayers with different possible configurations of Pd and Pt spins on honeycomb lattice, including paramagnetic (PM), ferromagnetic (FM), N$\acute{e}$el antiferromagnetic (NAFM), stripe AFM (SAFM), and zigzag AFM (ZAFM) configurations, as shown in Fig. ~\ref{fig3} (c). The results are summarized in Table ~\ref{tab:magnet}. One can observe that the out-of-plane FM (FM$^z$) state has the lowest energy among them. We further calculated the energies for FM configurations by rotating the magnetic direction deviated from the z-axis, and found that the FM$^z$ state is the most energetically favorable, which shows an Ising behavior of PdBr$_3$ monolayer.


Thus, the effective Hamiltonian can be described by $H_{spin}=-\sum_{\langle i,j\rangle}JS_{i}^{z}S_{j}^{z}$  where $J$ represents the nearest-neighbor exchange integral, $S_{i,j}^{z}$ is the spin operator, and ${\langle i,j\rangle}$ denotes the summation over nearest neighbors. $J$ can be determined by the difference of energies between ZAFM$^z$ and FM$^z$ configurations because the ZAFM$^z$ possesses the lowest energy among those AFM configurations (see Table ~\ref{tab:magnet}), which was estimated to be 79.2 meV and 84.9 meV for PdBr$_3$ and PtBr$_3$ monolayers, respectively.

\begin{table}[t]
	\caption{The total energy $E_\mathrm{tot}$ per unit cell for PdBr$_3$, PtBr$_3$, PdI$_3$ and PtI$_3$ monolayers (in meV, relative to $E_\mathrm{tot}$ of FM$^z$ (FM$^x$) ground state) for several spin configurations of Pd (Pt) atoms calculated by GGA+SOC+U method.}\label{tab:magnet}
	\begin{tabular}{cccccccc}
		\hline
		                     & FM$^z$ & NAFM$^z$ & SAFM$^z$ & ZAFM$^z$ &FM$^x$ &FM$^y$  &PM\\
		\hline \hline
		PdBr$_3$      &0.0     &141.0     &119.8     &79.2     &8.0     &7.7  &472.1\\
        PtBr$_3$      &0.0     &207.5     &163.2     &84.9     &1.8     &1.7  &313.5\\
		\hline
		                     & FM$^x$ & NAFM$^x$ & SAFM$^x$ & ZAFM$^x$ &FM$^y$ &FM$^z$  &PM\\
		\hline \hline
        PdI$_3$       &0.0     &110.1     &93.7     &47.9     &0.0     &13.8  &1306\\
        PtI$_3$   &0.0     &96.8     &67.2     &52.4     &0.0     &5.4  &260.7\\
		\hline
	\end{tabular}
\end{table}

Monte Carlo (MC) simulations \cite{Wolff1989} based on Ising model were carried out to calculate the Curie temperature. The MC simulation was performed on 60$\times$60 2D honeycomb lattice using 10$^6$ steps for each temperature. The magnetic moment as a function of temperature is shown in Fig. ~\ref{fig3} (d). It can be seen that the magnetic moment decreases rapidly to vanish at about 350 K and 375 K for PdBr$_{3}$ and PtBr$_{3}$ monolayers, respectively, indicating that PdBr$_{3}$ and PtBr$_{3}$ monolayers can be potential candidates of ferromagnetic semiconductors at room temperature.

\begin{figure}[!hbt]
  \centering
  \includegraphics[scale=0.33,angle=0]{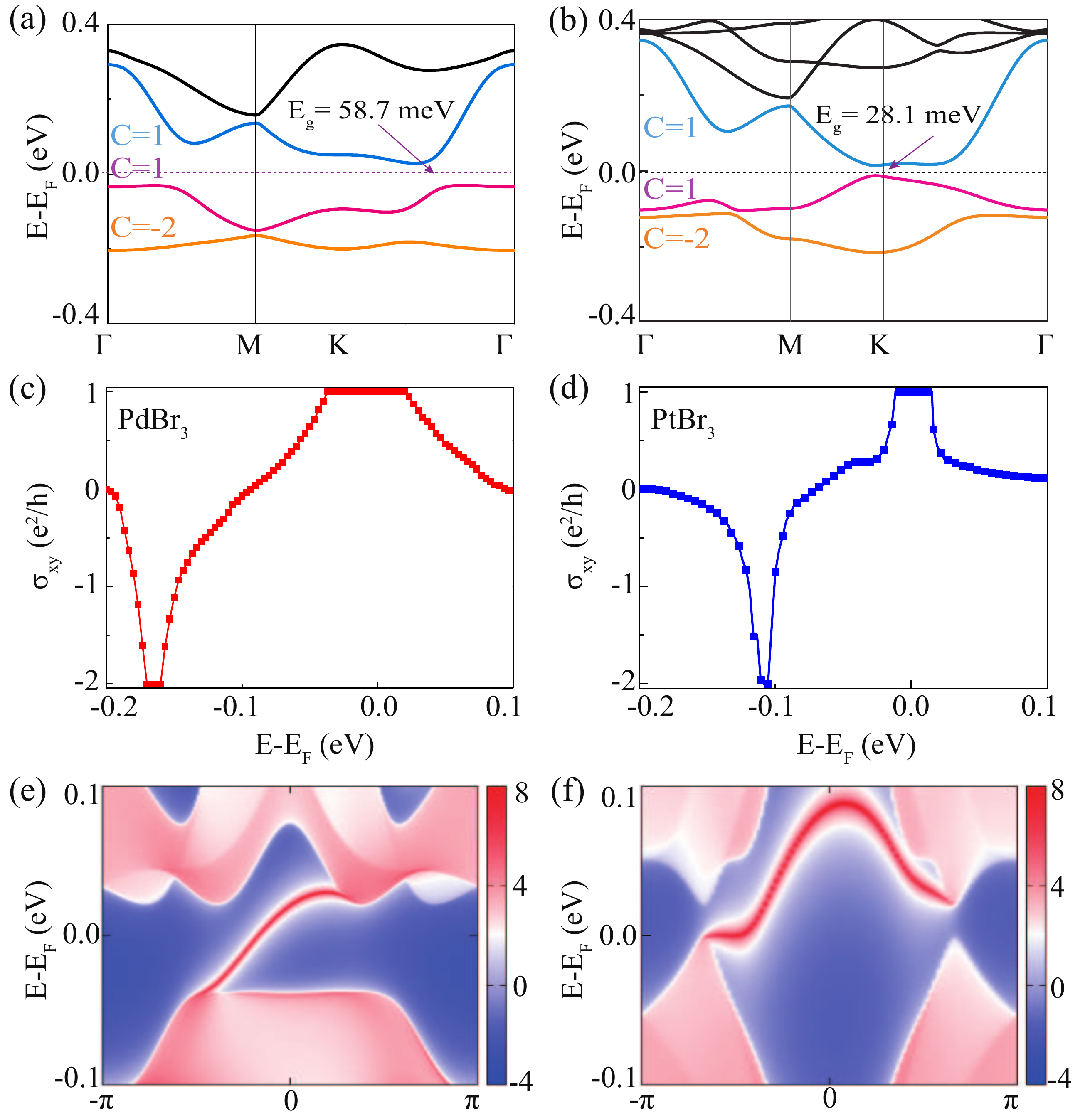}\\
  \caption{{\bfseries Room temperature QAHE in PdBr$_3$ and PtBr$_3$ monolayers.} The band structure of (a) PdBr$_3$ and (b) PtBr$_3$ monolayers with SOC, where Chern number $C$ of nontrivial bands near Fermi level is indicated, where the band gaps are 58.7 meV and 28.1 meV for PdBr3 and PtBr3 monolayers, respectively. Anomalous Hall conductivity of (c) PdBr$_3$ and (d) PtBr$_3$ monolayers as a function of energy near Fermi level. The surface states of (e) PdBr$_3$ and (f) PtBr$_3$ monolayers. Warmer colors represent higher local DOS, and blue regions indicate the bulk band gap. Results are obtained by the GGA+SOC+U calculations.}\label{fig4}
\end{figure}


In transition metal compounds, the $d$-orbitals are usually not fully filled, and the Coulomb correlation $U$ cannot be ignored. Although the accurate values of $U$ for PdBr$_{3}$ and PtBr$_{3}$ are not known, the correlation interaction $U$ for 4d electrons of Pd atoms in PdBr$_{3}$ monolayer should be smaller than 3 eV of the impurity Pd \cite{Solovyev1994}. The correlation interaction $U$ for 5$d$ electrons is typically weak and the reasonable value is usually less than 1 eV. Thus, we take $U$ = 2.5 eV for PdBr$_{3}$ and $U$ = 0.5 eV for PtBr$_{3}$ in calculations. The electronic band structures of PdBr$_{3}$ and PtBr$_{3}$ monolayers based on GGA+SOC+$U$ were plotted in Figs. \ref{fig4} (a) and \ref{fig4} (b), respectively. The SOC opens band gaps $E_{g}$ = 58.7 meV and 28.1 meV for PdBr$_{3}$ and PtBr$_{3}$ monolayers, respectively, and $E_{g}/k_{B}$ are higher than room temperature. Because GGA usually underestimates the band gaps, HSE06 was also employed to check the band gaps. Our calculations show that the band gaps become 100.8 meV and 45 meV with HSE06 for PdBr$_3$ and PtBr$_3$ monolayers, respectively.

To investigate the topological properties of PdBr$_{3}$ and PtBr$_{3}$ monolayers, we first calculated the gauge invariant Berry curvature $\Omega_{z}(K)$ in momentum space. Topologically nontrivial band structure is characterized by a nonzero Chern number $C$ that counts the number of edge states. The Chern number $C$ obtained by integrating the Berry curvature $\Omega_{z}(K)$ over the BZ is 1. As expected from a nonzero Chern number, the anomalous Hall conductivity shows a quantized charge Hall plateau of $\sigma_{xy}=Ce^{2}/h=e^{2}/h$ as shown in Figs. ~\ref{fig4} (c) and ~\ref{fig4} (d).

The Fermi surface of 2D materials can be adjusted by using a gate voltage. It is intriguing to find when the occupancy of electrons decreases by one, the Chern number becomes -2, while the occupancy increases by one, the Chern number is still 1. When the $e_{g}$ bands are half occupied ($E-E_{F}\approx0.35$ eV) or non-occupied ($E-E_{F}\approx-0.25$ eV), the Chern numbers become 0. Thus, the anomalous Hall conductivity can be tuned by a slight shift of Fermi level for PdBr$_{3}$ and PtBr$_3$ monolayers, as shown in Figs. ~\ref{fig4} (c) and ~\ref{fig4} (d).

According to the bulk-edge correspondence \cite{Hatsugai1993}, the nonzero Chern number is closely related to the number of nontrival chiral edge states that emerge inside the bulk gap of a semi-infinite system. As presented in Figs. \ref{fig4} (e) and \ref{fig4} (f), there is one gapless chiral edge state connecting the valence and conduction bands.

In order to verify the stability of QAHE against the in-plane strain effect, we also calculated the band structure and Hall conductance of PdBr$_{3}$ and PtBr$_{3}$ monolayers at different lattice constants from 5$\%$ tensile strain to 5$\%$ compression, and found that in these cases the band gap is only changed slightly but not closed, indicating that the QAHE in PdBr$_{3}$ and PtBr$_{3}$ monolayers is robust against the strain effect.

{\color{blue}{\em Discussions}}---To inspect how robust is the Ising behavior of PdBr$_3$ and PtBr$_3$ monolayers, with the same method, we have calculated the magnetic anisotropy of PtBr$_3$, PdBr$_3$, and CrI$_3$ monolayers, where the latter has been realized in recent experiment. The energy difference between the in-plane and out-of-plane ferromagnetic configurations is about 3.6 eV per unit cell for CrI$_3$ monolayer~\cite{Griffiths1964,Huang2017a,Samarth2017}, while it is 1.7 eV for PtBr$_3$ monolayer and 7.7 eV for PdBr3 monolayer. The CrI$_3$ monolayer has been verified to be an Ising-type ferromagnetism with out-of-plane magnetization. The PdBr$_3$ monolayer should be an Ising-type ferromagnetism with stronger anisotropy than CrI$_3$. To study the PtBr$_3$ monolayer, we take the Hamiltonian given in Ref.~\cite{Xu2018}, where both Kitaev interaction and single-ion anisotropy (SIA) were considered, by which the Ising behavior of CrI$_3$~\cite{Griffiths1964,Huang2017a,Samarth2017} and Heisenberg behavior of CrGeTe$_3$~\cite{Mermin1966,Xing2017,Gong2017,Samarth2017} can be interpreted. Our results show that, for PtBr$_3$ monolayer, both Kitaev anisotropy and SIA are numerically found to be negative, which determines the Ising-type behavior of PtBr$_3$ monolayer with out-of-plane magnetization.

To check the influence of the Coulomb interaction $U$ on the band gaps, we have calculated the bands using GGA+SOC+$U$ with different values of $U$ for PdBr$_3$ and PtBr$_3$ monolayers. The results show that the band gap changes from 61.3 meV to 56.9 meV with $U$ in the range of 2 eV to 3 eV for PdBr$_3$ monolayer, and the band gap changes from 28.1 meV to 33.7 meV with $U$ in the range of 0.5 eV to 1 eV for PtBr$_3$ monolayer. In all the cases, the semiconducting state is maintained. Note that the typical values of $U$ for 4d and 5d transition metals are adopted in our paper. It is known that the GGA-type calculation usually underestimates the band gap, and the hybrid functional (such as HSE06) calculation can give values approaching the experimental result. By HSE06 method, we found that the band gaps become 100.8 meV and 45 meV for PdBr$_3$ and PtBr$_3$ monolayers, respectively. As a result, the semiconducting states for PdBr$_3$ and PtBr$_3$ monolayers are robust.

The band gap is opened due to the SOC. It is known that the SOC in Pd(4d) atom is much smaller than that in Pt(5d) atom. However, one may see a larger band gap for PdBr$_{3}$ [Fig. \ref{fig4} (a)], and a smaller band gap for PtBr$_{3}$ [Fig. ~\ref{fig4} (b)]. To understand this puzzle, we studied the on-site Coulomb interaction $U$ dependence of the band gap for PdBr$_{3}$ and PtBr$_{3}$ monolayers by the GGA+SOC+$U$ calculations. The band gap, which originates from SOC, increases with the on-site Coulomb interaction $U$. Consider the Coulomb interaction between the orbitals with the same spin $\sigma$, which is given by $H_{l^{z}\sigma}=(U'-J_{H})(\overline{n}^2-\frac{1}{4}\lambda_{SO}^2(l^{z})^2\sigma^2(\delta n)^2)$, where $U(U')$ is the on-site Coulomb repulsion within (between) the orbitals, and $J_{H}$ is the Hund coupling between the orbitals. The relationship $U=U'+2J_{H}$~\cite{Maekawa2004} holds in the atomic limit. To compensate the interaction, the band splitting due to SOC will increase with increasing $U$. In fact, the enhancement (renormalization effect) of SOC due to Coulomb interaction $U$ was also discussed in the spin Hall effect in Au metal with Fe impurity \cite{Gu2010}. The increased $U$ will produce the enhanced SOC, which will induce the enhanced band gap. It turns out that the larger band gap and larger anisotropy in 4d compound PdBr$_{3}$ mainly comes from the larger renormalization effect of SOC of Pd atom with a larger $U$ value in PdBr$_{3}$.
The calculation on the $U$ dependence of Curie temperature shows that when the Coulomb interaction of PdBr$_{3}$ monolayer ranges from 1 eV to 3 eV, the Curie temperature changes from 310 K to 383 K, and when the Coulomb interaction of PtBr$_{3}$ monolayer ranges from 0 to 1 eV, the Curie temperature changes from 340 K to 400 K, which is pretty higher than room temperature, while in both cases the QAHE preserves.

The PdI$_3$ and PtI$_3$ monolayers were also calculated using the same method as the above two monolayers. There are no imaginary frequencies for PdI$_3$ and PtI$_3$ monolayers, indicating that they are also dynamically stable. The stability of these two monolayers were also checked by the formation energy (Table~\ref{tab:formation}) and free energy (Fig.~\ref{fig2}). The PdI$_3$ and PtI$_3$ monolayers both show the in-plane magnetization, and the Curie temperatures can be estimated~\cite{Spirin2003} to be T$_c$ = 150 K and 164 K, respectively, which is higher than the values of T$_c$ in the present experiments.

Based on the 2D room temperature magnetic semiconductors with QAHE, new spintronic devices can be designed. For example, by magnetic and spin-orbit proximity effects in bilayer junctions, it enables us to design new functional spintronic devices and to overcome various limitations~\cite{Zutic2019}.

{\color{blue}{\em Conclusions}}---Based on first-principles calculations, we report that PdBr$_3$ and PtBr$_3$ monolayers are room temperature out-of-plane ferromagnetic semiconductors with Curie temperatures 350 K and 375 K, respectively. Room temperature QAHE can also be implemented in 2D PdBr$_3$ and PtBr$_3$ with energy band gaps of 58.7 meV and 28.1 meV with GGA and 100.8 meV and 45 meV with HSE06, respectively. The large band gaps come from multi-orbital electron correlations. Here we would like to mention that the bulk PtBr$_3$ has been synthesized long time ago. By paying some efforts, the experimental realization of 2D PtBr$_3$ crystals may also be possible. Therefore, we expect that the 2D room-temperature ferromagnetic semiconductors PdBr$_3$ and PtBr$_3$ with QAHE predicted in the present work can be implemented in experiments in near future. By applying the room temperature ferromagnetic semiconductors with QAHE, the progress in developing spintronic devices and dissipationless devices is highly expected.
\ \
\par
\ \

{\color{blue}{\em Acknowledgements}}--- BG is supported by NSFC (Grant No. Y81Z01A1A9), CAS (Grant No. Y929013EA2)   and UCAS (Grant No.110200M208). GS is supported in part by the the National Key R$\&$D Program of China (Grant No. 2018FYA0305800), the Strategic Priority Research Program of CAS (Grant Nos. XDB28000000, XBD07010100), the NSFC (Grant No. 11834014), and Beijing Municipal Science and Technology Commission (Grant No. Z118100004218001).

\end{document}